\documentclass[a4paper,12pt]{myarticle}
\usepackage{graphicx}
\renewcommand\figurename{\small {\sc Fig.}}
\renewcommand\tablename{\small {\sc Table}}
\begin{document}
\thispagestyle{empty}
\ \vspace{0cm}
\begin{center}
\renewcommand{\baselinestretch}{1.1} \normalsize 
{\bf \large Noisy Covariance Matrices and Portfolio Optimization} \par \vspace{0.7cm}
{\underline{Szil\'ard Pafka}$^{1}$ and Imre Kondor$^{1,2}$} \par \vspace{0.35cm}
{\it \small $^1$Department of Physics of Complex Systems, E\"otv\"os University \\
P\'azm\'any P.\ s\'et\'any 1/a, H--1117 Budapest, Hungary} \par \vspace{0.1cm}
{\it \small $^2$Market Risk Research Department, Raiffeisen Bank \\
Akad\'emia u.\ 6, H--1054 Budapest, Hungary} \par \vspace{0.45cm}
November, 2001 \par \vspace{1.1cm}
{\bf Abstract} \par \vspace{0.8cm}
\parbox{13cm}{\small
According to recent findings \cite{bouchaud,stanley}, empirical covariance 
matrices deduced from financial return series contain such a high amount of 
noise that, apart from a few large eigenvalues and the corresponding 
eigenvectors, their structure can essentially be regarded as random. 
In \cite{bouchaud}, e.g., it is reported that about 94\% of the spectrum 
of these matrices can be fitted by that of a random matrix drawn from 
an appropriately chosen ensemble. 
In view of the fundamental role of covariance matrices in the theory of 
portfolio optimization as well as in industry-wide risk management practices,
we analyze the possible implications of this effect.
Simulation experiments with matrices having a structure such as described in 
\cite{bouchaud,stanley} lead us to the conclusion that in the context 
of the classical portfolio problem (minimizing the portfolio variance 
under linear constraints) noise has relatively little effect. 
To leading order the solutions are determined by the stable, large 
eigenvalues, and the displacement of the solution (measured in variance) due 
to noise is rather small: depending on the size of the portfolio and on the 
length of the time series, it is of the order of 5 to 15\%. 
The picture is completely different, however, if we attempt to minimize 
the variance under non-linear constraints, like those that arise 
e.g.\ in the problem of margin accounts or in international capital adequacy regulation. 
In these problems the presence of noise leads to a serious instability and a 
high degree of degeneracy of the solutions.
\par \vspace{0.3cm} {\it Keywords:} noisy covariance matrices,
random matrix theory, portfolio optimization, risk management
} \end{center} \vspace{1cm} \par
\rule{5cm}{0.4pt} \par
{\small {\it E-mail:} syl@complex.elte.hu (S.\ Pafka),
ikondor@raiffeisen.hu (I.\ Kondor)
\newpage \setcounter{page}{1} 
\renewcommand{\baselinestretch}{1.1} \normalsize 


\section{Introduction}

The concept of financial risk, which attempts to quantify the uncertainty of the
outcome of an investment and hence the magnitude of possible losses, plays a fundamental role in finance today.
Portfolio optimization aims at giving a recipe for the composition of portfolios such that the overall risk is minimized for
a given reward, or, conversely, reward is maximized for a given risk.
For example, the classical portfolio optimization problem formulated first 
by Markowitz \cite{markowitz} relies on the variance as a risk measure and 
expected return as a measure for reward. 
Since the return on a portfolio is a linear combination of the returns on 
the assets forming the portfolio with weights given by the proportion of 
wealth invested in the assets, the portfolio variance can be 
expressed as a quadratic form of these weights with the volatilities and correlations as coefficients. 
For any practical use of the theory, it will, therefore, be necessary to
have reliable estimates for the volatilities and correlations, which, in most cases, are obtained from historical return series.
Actually, volatility and correlation estimates extracted from 
historical data have become standard tools also for several other risk 
management practices widely utilized in the financial industry.

Recently it has, however, been found by two independent groups \cite{bouchaud,stanley} that empirical covariance 
matrices deduced from financial return series contain such a high amount of 
noise that, apart from a few large eigenvalues and the corresponding 
eigenvectors, their structure can essentially be regarded as random. 
In \cite{bouchaud}, e.g., it is reported that about 94\% of the 
spectrum of correlation matrices determined from return series on 
the S\&P 500 stocks can be fitted by that of a random matrix drawn from an 
appropriately chosen ensemble. 
In view of these striking results, the Markowitz portfolio optimization scheme
based on a purely historical determination of the covariance matrix would
seem to be totally inadequate \cite{bouchaud,bouchaud-riskmagaz}, but 
the credibility of a number of standard risk management methodologies 
would also be shaken.

In this paper we will argue, however, that the impact of the results
of \cite{bouchaud,stanley} on the portfolio optimization problem may
not be as dramatic as one might have expected.
More specifically, it will be shown in a simulation example that for
parameter values typically encountered in practice, the risk of the 
optimal portfolio determined in the presence of noise is usually no more 
than 5--15\% higher than the risk without noise. 
Despite the high degree of noise of the covariance
matrix, which translates indeed into a significant displacement in 
the weights of the optimal portfolio, the effect on the actual risk at 
the optimum is only of second order and therefore less pronounced.
This suggests that some of the risk management methodologies based on 
empirical covariance matrices can actually be sufficiently accurate.
The main purpose of this paper is to point out that the results 
of \cite{bouchaud,stanley} and the practical usefulness of covariance 
matrices can, in fact, be reconciled.


\section{Results and Discussion}

We consider the following simplified version of the classical portfolio optimization
problem: the portfolio variance $\sum_{i,j=1}^n w_i\,\sigma_{ij}\,w_j$ 
is to be minimized under the budget constraint $\sum_{i=1}^n w_i=1$, where
$w_i$ denotes the weight of asset $i$ in the portfolio while $\sigma_{ij}$ 
represents the covariance matrix of returns (considered here as given). This means we exclude the riskless bond and seek the minimal risk portfolio in the space of risky assets. One might, of course, impose additional constraints (e.g. the usual one on the return), but the simplified form at hand provides the most convenient laboratory to test the effect of noise. The solution to the optimization problem can then be found
using the method of Lagrange multipliers, and after some trivial algebra one obtains for the weights of the optimal portfolio:
\begin{equation}
\label{eq:sol}
w_i^*=\frac{\sum_{j=1}^n \sigma_{ij}^{-1}}{\sum_{j,k=1}^n \sigma_{jk}^{-1}}.
\end{equation}

According to \cite{bouchaud,stanley}, correlation matrices determined from 
financial return series are such that apart from a few large eigenvalues 
and the corresponding eigenvectors, their structure is essentially random. Random matrix theory (RMT) allows one to calculate different
eigenvalue and eigenvector statistics e.g.\ of a matrix 
$C_{ij}=\frac{1}{T} \sum_{t=1}^{T} x_{it}\,x_{jt}$ determined from 
series of random variables $x_{it}$ independent and identically
distributed, of mean zero 
and of unit variance ($i=1,2,\ldots,n$ and $t=1,2,\ldots,T$),
see \cite{bouchaud,stanley} and references therein. 
The observed deviations of empirical correlation matrices from RMT predictions 
\cite{bouchaud,stanley,stanley-condmat,drozdz} are due to genuine
correlations between the financial series, while the apparently 
dominating random part can
be interpreted as noise superimposed on these correlations.
Therefore, any procedure using as input correlation matrices determined from
financial return series will be biased by a significant amount of noise, and for the practical applicability of the procedure it would
be highly desirable to know the magnitude of this bias.

In order to get an idea about the magnitude of the effect, we compare the solution obtained for a given noiseless covariance matrix
$\sigma^{(0)}_{ij}$ with that obtained when noise is added (we call
the new covariance matrix $\sigma_{ij}$). If, for example, 
$\sigma^{(0)}_{ij}$ is simply chosen to be an $n\times n$ identity matrix, noisy covariance
matrices $\sigma_{ij}$ can be generated as
\begin{equation}
\sigma_{ij}=\frac{1}{T} \sum_{t=1}^{T} x_{it}\,x_{jt},
\end{equation}
where $x_{it}\sim \textrm{i.i.d.\ N}(0,1)$.
(Of course, in the limit $T\to\infty$ the noise disappears and
$\sigma_{ij}\to\sigma_{ij}^{(0)}$.)
The solutions to the optimization problem in this setup are
$w_i^{(0)*}=\frac{1}{n}$ and $w_i^*$ given by Eq.\ (\ref{eq:sol}),
respectively.
The difference between the two solutions shows the displacement of the 
optimal portfolio due to noise and provides a measure for the effect of
noise on the optimization problem. 

\begin{figure}[!ht]
\begin{center}
\includegraphics[scale=0.5,angle=-90]{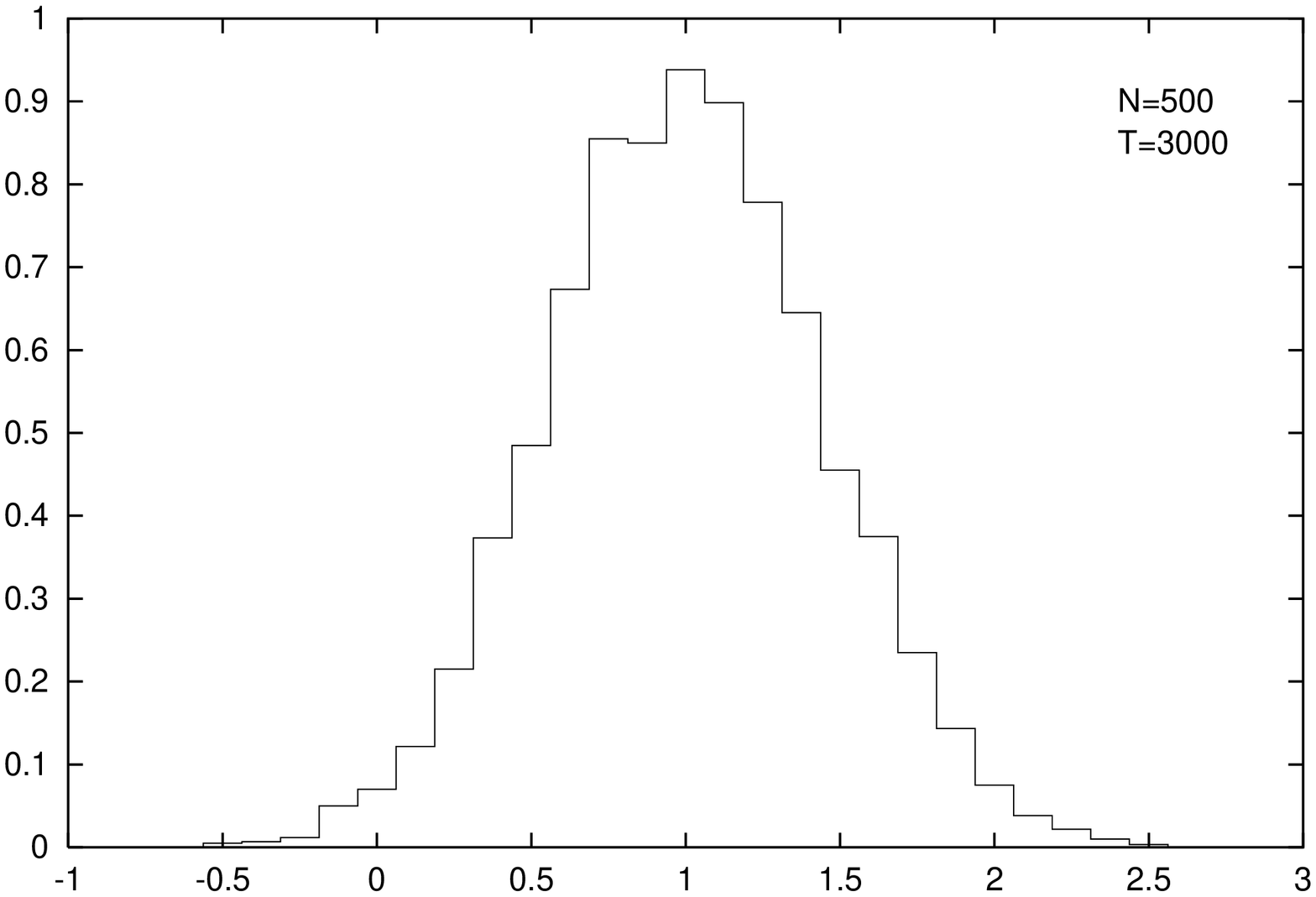}
\includegraphics[scale=0.5,angle=-90]{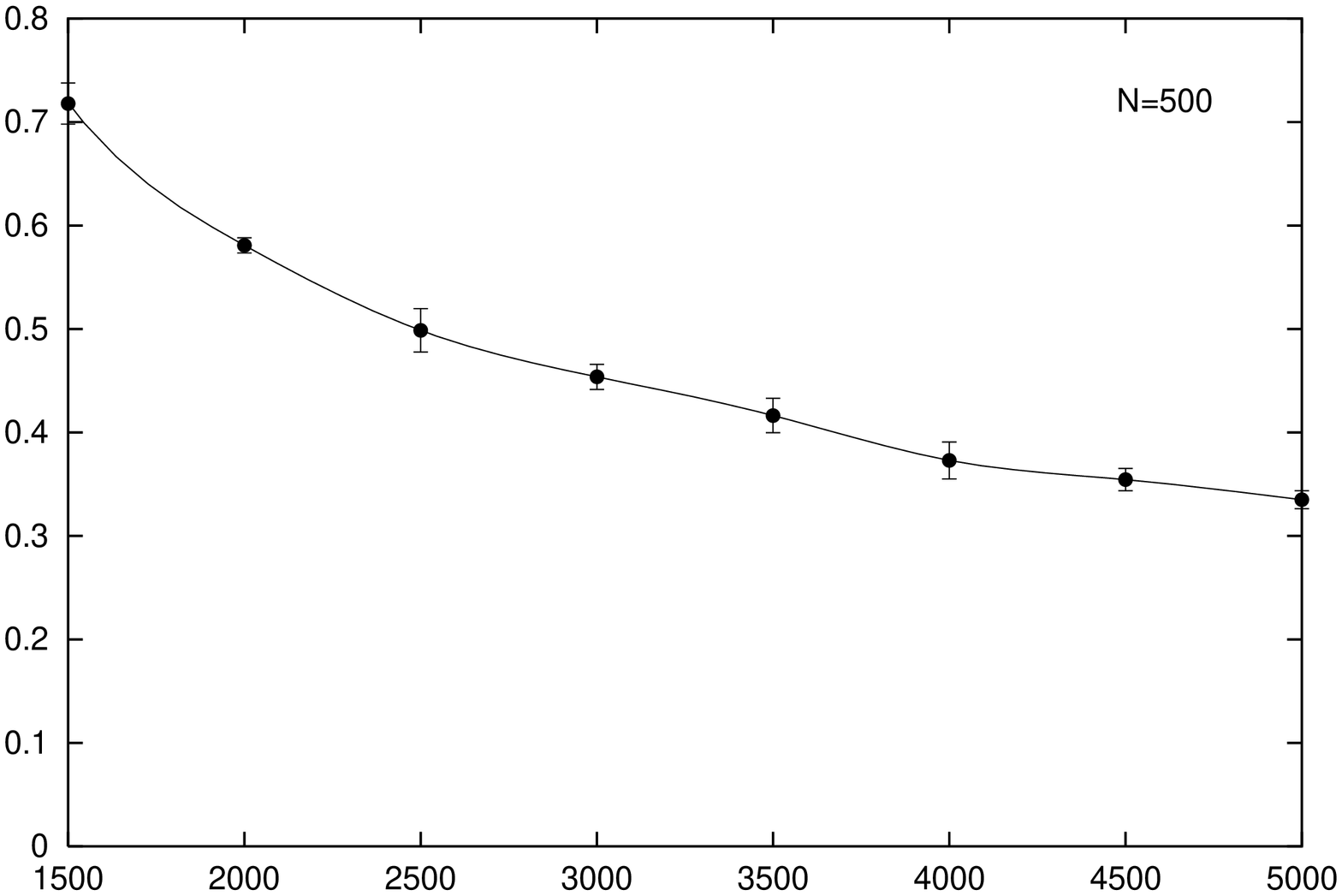}
\end{center}
\caption{({\it a}) Histogram of $n w_i^*$ illustrating the displacement of 
the solution in the presence of noise for $N=500$ and $T=3000$ (compare 
to $n w_i^{(0)*}=1$). To achieve smoothness, the histogram has been averaged 
over 10 samples for $\sigma_{ij}$.
({\it b}) Standard deviation of $n w_i^*$ for $N=500$ and 
different values for $T$.
\label{fig:w}}
\end{figure}

We have studied the behavior of $n w_i^*$ for different system
sizes $N$ and different time series lengths $T$. Without noise
this quantity is $n w_i^{(0)*}=1$ for any $i$. In the presence of noise, 
however, $n w_i^*$ oscillates around 1. The distribution of $n w_i^*$ 
for typical values of $N$ and $T$ is given in Fig.\ \ref{fig:w}({\it a}). 
It can be seen from the figure that weights as low as 0 or as high 
as 2 appear with non-negligible frequency, which suggests that 
the effect of
noise is quite strong, i.e.\ the optimal portfolio
obtained using the noisy covariance matrix may be rather different from
the ``true'' optimal portfolio.
The standard deviation of this distribution as a function of $T$ is
given in Fig.\ \ref{fig:w}({\it b}). One can see that the deviation 
from the optimal portfolio remains significant even for quite large
$T$.

However, the effect of noise should
be assessed not so much on the basis of the change in composition, but rather of the shift in the volatility of the optimal
portfolio, because this is the only factor rational investors should 
actually care about in our simplified optimization problem.
Let us, therefore, compare the variance of the ``true'' (noiseless) optimal
portfolio ${\sigma_p^{(0)}}^2=\sum_{i,j=1}^n w_i^{(0)*}\,\sigma_{ij}^{(0)}\,
w_j^{(0)*}=\sum_{i=1}^n {w_i^{(0)*}}^2$ with the ``true'' variance
of the optimal portfolio obtained in the presence of noise
$\sigma_p^2=\sum_{i,j=1}^n w_i^{*}\,\sigma_{ij}^{(0)}\,
w_j^{*}=\sum_{i=1}^n {w_i^{*}}^2$. More precisely, we have calculated the
volatility ratio $q=\frac{\sigma_p}{\sigma_p^{(0)}}$ that measures
the increase in volatility (and therefore decrease in efficiency) of the
optimal portfolio due to noise. Fig.\ \ref{fig:q} shows the magnitude
of this quantity as a function of $T$ for $N$ given. It can be seen
from the figure that for $N=500$ and $T>3000$ the increase in volatility
due to noise is less than 10\%. The decrease in efficiency 
is, in most cases, of the order of 5--15\% that seems reasonable from
a practical point of view ($2000<T<5000$).
It seems, therefore, that despite its obvious effect on the weights of 
the optimal portfolio, noise has a significantly less pronounced impact 
on the risk at the optimum.

\begin{figure}[h!]
\begin{center}
\includegraphics[scale=0.5,angle=-90]{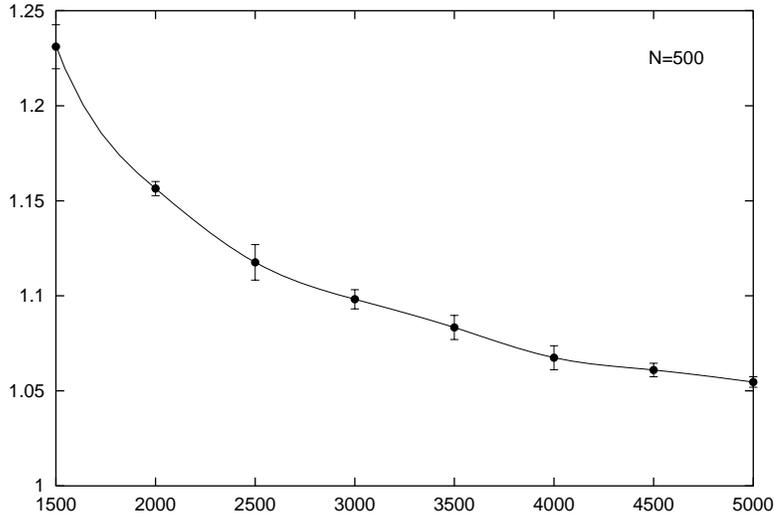}
\end{center}
\caption{Ratio $q=\frac{\sigma_p}{\sigma_p^{(0)}}$ quantifying the 
decrease in efficiency of the optimal portfolio due to noise in the
covariance matrix as a function of $T$ ($N=500$).
\label{fig:q}}
\end{figure}

It is interesting to estimate the magnitude of this effect
for values of $N$ and $T$ 
similar to those considered in \cite{bouchaud,stanley}. In \cite{bouchaud}
daily returns on $N=406$ stocks of the S\&P 500 over the period
1991--1996 (a total of $T=1309$ daily observations) have been used
for constructing the correlation matrix. We have found that for this
portfolio size and for this time series length the value of $q$ is around 1.20, i.e.\
the risk of the optimal portfolio in the presence of 
noise is about 20\% larger than without.
As for \cite{stanley}, in this paper 30-minute returns on the largest
$N=1000$ U.S.\ stocks over the two-year period 1994--1995 (a total
of $T=6448$ data points for each series) have been used, for which 
$q$ is about 1.09, i.e.\ the decrease in efficiency due to noise is
only around 9\%.
According to these findings, the impact of noise on the portfolio
optimization problem does not seem to be as dramatic as one might have feared and that, despite the high level of noise, empirical covariance matrices
can still be used as input
for a portfolio optimization problem without loosing too much of accuracy.

Moreover, for smaller system sizes $N$ we need shorter time series
lengths $T$ to achieve the same degree of precision. For example, for
$N=100$ and $T=500$ the increase in volatility is 11\%, while for 
$T=1000$ it is only about 5\%. Therefore, if one uses covariance 
matrices obtained e.g.\ from 4 years of daily returns (around 
1000 observations) say 
on the S\&P 100 stocks, the loss in efficiency due to noise in the 
covariance matrix is only about 5\%.

In order to give a better representation of the actual structure of empirical correlation matrices, we have repeated our experiments 
with matrices which, in addition to the pure random part given by
Eq.\ (\ref{eq:sol}), have one clearly distinct eigenvalue chosen to be 
about 25 times larger than the largest eigenvalue predicted by RMT
(see \cite{bouchaud,stanley}), with a corresponding eigenvector in the
direction of $(1,1,\ldots,1)$. The displacement of the risk associated
with the optimal portfolio due to noise has been found to be of the same
order of magnitude as in the cases discussed earlier. 

The explanation for the lack of a more dramatic effect on the portfolio in the presence of noise is actually very simple. A function $f(x)$ with 
a single well-defined flat minimum at $x^*$ (e.g.\ a quadratic function) varies 
slowly in the neighbourhood of the minimum, therefore, the value $f(x)$ need
not be much higher than $f(x^*)$, even for 
significant deviations of $x$ from the minimum. 
In our case, the volatility of the portfolio is exactly such a function
of the weights $w_i$, and therefore even if the 
weights deviate 
significantly from the optimal ones (as they do), the risk of the portfolio will not be dramatically affected. 
Since rational investors should not care about the composition of their
portfolios but only about its risk, the effect of noisy covariance
matrices on the portfolio optimization problem will be less significant than expected.
In other words, it appears that despite noise, covariance matrices deduced
from financial return series can have, in certain cases, 
a reasonable practical use.

Let us note, however, that the picture becomes 
completely different if we consider an
optimization problem with non-linear constraints like those that
arise e.g.\ in the case of margin accounts or in international capital adequacy 
regulation \cite{spin-bouchaud, spin-kondor, kondor}. In these cases the budget 
constraint has the form 
$\sum_{i=1}^n \gamma_i\,|w_i|=1$. As shown in \cite{spin-bouchaud}, this
problem maps exactly onto
finding the ground states of a long-range spin glass, and now the presence 
of noise leads to a serious instability and a high degree of degeneracy 
of the solutions.
The results of \cite{spin-bouchaud} have far reaching economic implications, 
the analysis of which are, however, far beyond 
the scope of this short note.


\section{Conclusion}

In this paper the impact of noisy covariance matrices on the portfolio
optimization problem has been investigated. Earlier studies 
\cite{bouchaud,stanley} have pointed out that a large part of the
spectrum of empirical covariance matrices deduced from financial return 
series corresponds to that of a purely random matrix. The presence
of such a high level of noise in covariance data
could have had devastating consequences for the 
reliability of different risk management practices based on the use of 
these matrices.
We have analyzed the impact of this noise on the classical portfolio 
optimization problem and found that the risk of the optimal portfolio 
determined in the presence of noise is typically no more than 5--15\% 
higher than in the absence of it, showing that the decrease in
efficiency of the optimal portfolio is actually much less dramatic.
This suggests that the important results 
of \cite{bouchaud,stanley} and the practical usefulness of covariance 
matrices can, in fact, be reconciled.


\section*{Acknowledgements}

This work has been supported by the Hungarian National Science
Found OTKA, Grant No.\ T 034835.




\begin{thebibliography}{99}

\bibitem{bouchaud}
L.\ Laloux, P.\ Cizeau, J.-P.\ Bouchaud and M.\ Potters, 
Phys.\ Rev.\ Lett.\ {\bf 83}, 1467 (1999).

\bibitem{stanley}
V.\ Plerou, P.\ Gopikrishnan, B.\ Rosenow, L.\ A.\ N.\ Amaral and 
H.\ E.\ Stanley, 
Phys.\ Rev.\ Lett.\ {\bf 83}, 1471 (1999).

\bibitem{markowitz}
H.\ Markowitz, {\it Portfolio Selection: Efficient Diversification
of Investments} (J.\ Wiley and Sons, New York, 1959).

\bibitem{bouchaud-riskmagaz}
L.\ Laloux, P.\ Cizeau, J.-P.\ Bouchaud and M.\ Potters, 
Risk {\bf 12}, No.\ 3, 69 (1999).

\bibitem{stanley-condmat}
V.\ Plerou, P.\ Gopikrishnan, B.\ Rosenow, L.\ A.\ N.\ Amaral, 
T.\ Guhr and H.\ E.\ Stanley, 
e-print cond-mat/0108023.

\bibitem{drozdz}
S.\ Dro\.zd\.z, F.\ Gr\"ummer, F.\ Ruf and J.\ Speth,
e-print cond-mat/0103606.

\bibitem{spin-bouchaud}
S.\ Galluccio, J.-P.\ Bouchaud and M.\ Potters,
Physica A {\bf 259}, 449 (1998).

\bibitem{spin-kondor}
A.\ G\'abor and I.\ Kondor,
Physica A {\bf 274}, 222 (1999).

\bibitem{kondor}

I.\ Kondor, Int.\ J.\ of Theor.\ and Appl.\ Finance 
{\bf 3}, 537 (2000).

\end{thebibliography}
\end{document}